# Facet-dependent giant spin orbit torque in single crystalline antiferromagnetic Ir-Mn / ferromagnetic permalloy bilayers

Weifeng Zhang[1,2*], Wei Han[1,3*], See-hun Yang[1], Yan Sun[4], Yang Zhang[4], Binghai Yan[4], and Stuart S. P. Parkin[1,5§]

[1]IBM Research - Almaden, San Jose, California 95120, USA

[2]Department of Material Science Engineering, Stanford University, Stanford, California 94305, USA

[3]International Center for Quantum Materials, School of Physics, Peking University, Beijing 100871, China

[4]Max Planck Institute for Chemical Physics of Solids, Dresden D-01187, Germany

[5]Max Planck Institute for Microstructure Physics, Halle (Saale) D-06120, Germany

*These authors contributed equally to the work

§ Email: stuart.parkin@mpi-halle.mpg.de (S.S.P.P.)

There has been considerable interest in spin-orbit torques for the purpose of manipulating the magnetization of ferromagnetic (FM) films or nano-elements for spintronic technologies[1-6]. Spin-orbit torques are derived from spin currents created from charge currents in materials with significant spin-orbit coupling that diffuse into an adjacent FM material[2,3,7,8]. There have been intensive efforts to search for candidate materials that exhibit large spin Hall angles, i.e. efficient charge to spin current conversion[2,5,7,9-12]. Here we report, using spin torque ferromagnetic resonance, the observation of a giant spin Hall angle $\theta_{SH}^{eff}$ of up to ~0.35 in (100) oriented single crystalline antiferromagnetic (AF) IrMn$_3$ thin films, coupled to ferromagnetic permalloy layers, and a $\theta_{SH}^{eff}$ that is about three times smaller in (111) oriented films. For the (100) oriented samples we show that the magnitude of $\theta_{SH}^{eff}$ can be significantly changed by manipulating the populations of the various AF domains through field annealing. Using *ab-initio* calculations we show that the triangular AF structure of IrMn$_3$ gives rise to a substantial intrinsic spin Hall conductivity that is three times larger for the (100) than for the (111) orientations, consistent with our experimental findings.

Spin orbitronics is an emerging field of current great interest in which spin-orbit coupling gives rise to several novel physical phenomena, many of which also have potential technological significance[6]. Of particular interest is the spin Hall effect[13] in which conventional charge current densities $J_c$ are converted to pure spin current densities $\hbar/2J_{SH}$ via spin-dependent scattering within the material. The conversion efficiency is described by the spin Hall angle, $\theta_{SH} = eJ_{SH}/J_c$. A widely used technique for measuring the spin Hall angle - spin torque ferromagnetic resonance (ST-FMR) - is by comparing the anti-damping spin torque generated by an RF spin current that diffuses into an adjacent ferromagnetic layer with the field torque generated by an RF charge current. ST-FMR gives an effective spin Hall angle $\theta_{SH}^{eff}$ since there are other potential sources of spin current, as we discuss further below. The largest values of $\theta_{SH}^{eff}$ in conventional metals, that have been reported to date, include the heavy metals β-tantalum[2], and β-tungsten[11], platinum (once the interface transparency is taken into account)[9] and bismuth doped copper[10].

A prerequisite for large $\theta_{SH}$ is significant spin-orbit coupling. Of potential interest are compounds such as Ir$_{1-x}$Mn$_x$ which are antiferromagnetic for various ranges of x. Such compounds are widely used in spintronic devices due to the exchange bias fields that they imprint on neighboring ferromagnetic layers that leads to a unidirectional magnetic anisotropy[14]. Even more interestingly, recent theoretical studies predict that chemically ordered IrMn$_3$ should exhibit a large anomalous Hall effect (AHE) due to its unusual triangular magnetic structure[15,16]. Here we report the observation of giant $\theta_{SH}^{eff}$ in Ir-Mn films, and a large facet dependence of $\theta_{SH}^{eff}$ in IrMn$_3$, which we find can be dramatically enhanced by annealing in magnetic fields oriented perpendicular, but not parallel, to the film layers. Moreover, we show, with support from *ab-initio* calculations, that these results can be accounted for by a substantial intrinsic spin Hall effect that arises from the AF triangular structure of IrMn$_3$.

Microstrip devices, 10 μm wide and 100 μm long, are fabricated using standard optical lithography and ion milling techniques from Ir$_{1-x}$Mn$_x$ / Py (Py = permalloy) / TaN films. The films are deposited by magnetron sputtering (see Methods for details) on Si(100) substrates (covered with 25 nm SiO$_2$), (100) MgO, and (0001) Al$_2$O$_3$ substrates at room temperature (RT) in a high vacuum deposition chamber. ST-FMR measurements are carried out by passing a high frequency RF current into the device ($J_C$ is the charge current density in Fig. 1a) in the presence of an external field $H_{ext}$ that is applied in the film plane at 45 degrees to the microstrip. The charge current that flows in the Ir$_{1-x}$Mn$_x$ layer generates two distinct torques on the Py layer, namely a spin-orbit torque and an Oersted field torque. The time derivative of the magnetization of Py, $\hat{m}$, is given by [7,17]:

$$\frac{d\hat{m}}{dt} = -\gamma \hat{m} \times \vec{H}_{eff} + \alpha \hat{m} \times \frac{d\hat{m}}{dt} - \gamma \hat{m} \times \vec{H}_{rf} + \gamma \frac{\hbar}{2\mu_0 M_s t} J_S (\hat{m} \times \hat{\sigma} \times \hat{m}) \quad (1)$$

where $\gamma$ is the gyromagnetic ratio, $\vec{H}_{eff}$ is composed of the external field, the exchange bias field and the out-of-plane demagnetization field, $\alpha$ is the Gilbert damping constant, $\vec{H}_{rf}$ is the Oersted field, $\hbar/2J_S$ is the spin current density generated in the Ir$_x$Mn$_{1-x}$ layer, $\mu_0$ is the permeability in vacuum, $M_S$ is the saturation magnetization of the Py layer, which is measured using vibrating sample magnetometry on the un-patterned film prior to device fabrication, $t$ is the thickness of the Py layer, and $\hat{\sigma}$ is the direction of the injected spin moment. We note that $J_S$ represents any spin current that is generated by passage of the charge current through the device and that gives rise to an anti-damping torque. There may be, as we discuss below, mechanisms beyond spin Hall effects that generate spin currents in the device. The Oersted field depends on the RF current and the thickness of the Ir$_x$Mn$_{1-x}$ layer $d$, and can be expressed as $H_{rf} = \frac{1}{2}J_C d$ for our experiments in which the Ir$_{1-x}$Mn$_x$ films are comparatively thin (10-150 Å).

Mixing of the RF charge current through the device and the consequent oscillating magnetization of the Py layer that affects the device resistance through its anisotropic magnetoresistance generates a DC voltage, $V_{mix}(H_{ext})$. Fig. 1b shows typical ST-FMR spectra, i.e. $V_{mix}(H_{ext})$ as a function of $H_{ext}$, measured on ~60 Å polycrystalline IrMn$_3$ (p-IrMn$_3$) /60 Å Py for RF frequencies varying between 9 and 14 GHz. As the RF frequency increases, the resonance field, $H_{res}$, increases, as expected from the Kittel model, and the magnitude of $V_{mix}$ decreases. $V_{mix}(H_{ext})$ has two components, an anti-symmetric component due to the Oersted field generated torque, and a symmetric component that arises from the spin orbit torque. The ST-FMR curves are fitted with symmetric and anti-symmetric Lorentzian functions, according to[7,9]:

$$V_{mix} = c[V_S \frac{\Delta H^2}{\Delta H^2 + (H_{ext} - H_{res})^2} + V_A \frac{\Delta H(H_{ext} - H_{res})}{\Delta H^2 + (H_{ext} - H_{res})^2}] \qquad (2)$$

where $c$ is a constant, $V_S$ and $V_A$ are the amplitudes of the symmetric and anti-symmetric components, respectively, of the DC mixed voltage, and $\Delta H$ is the half line-width. $\theta_{SH}^{eff}$ is given by[7,9]:

$$\theta_{SH}^{eff} = \frac{eJ_S}{J_C} = \frac{V_S}{V_A} \frac{e\mu_0 M_S td}{\hbar} [1+(\frac{4\pi M_{eff}}{H_{ref}})]^{1/2} \quad (3)$$

where $M_{eff}$ is the effective magnetization that can be extracted by fitting the resonance frequency $f_{res}$ as a function of $H_{res}$ using the Kittel formula[18] $f_{res} = (\frac{\gamma}{2\pi})[(H_{res}+H_B)(H_{res}+H_B+4\pi M_{eff})]^{1/2}$, in which $H_B$ is the in-plane exchange-bias field due to the interaction between Py and the $IrMn_3$ layer. For polycrystalline $Ir_{1-x}Mn_x$ films, the measured in-plane exchange-bias field is very small compared to $H_{res}$ (Fig. 2b).

Typical results of fitting the experimental data in Fig. 1b with equation (2) are shown in Fig. 1c from which the fitting parameters, $V_S$, $V_A$, $H_{res}$ and $\Delta H$ can be obtained. Using these parameters $\theta_{SH}^{eff}$ is estimated to be 0.092±0.003 (at 11 GHz) based on equation (3). $\theta_{SH}^{eff}$ exhibits little variation as a function of the RF frequency, as shown in Fig. 1d. The dependence of the ST-FMR spectra on the thickness of the p-$IrMn_3$ layer, $d$, is shown in Fig. 1e. From fits to these data using equation (2), the magnitude of the $V_S$ varies little as a function of $d$, while $V_A$ increases approximately linearly with $d$ (see inset to Fig. 1e), consistent with larger Oersted field torques, as the RF charge current flowing in the p-$IrMn_3$ layer increases. As shown in Fig. 1f, $\theta_{SH}$ shows a weak dependence on $d$, which suggests that the spin diffusion length in $IrMn_3$ is small (<~10 Å).

The magnitude of $\theta_{SH}^{eff}$ in p-$IrMn_3$ is considerable (~0.1) and much larger than $\theta_{SH}^{eff}$ measured indirectly from spin pumping from permalloy into $Ir_{50}Mn_{50}$ layer[19] but comparable to that found in spin pumping studies from yttrium iron garnet into $Ir_{20}Mn_{80}$[20]. To explore its origin, a systematic study of various $Ir_{1-x}Mn_x$ concentrations (x=0, 0.17, 0.53, 0.64, 0.75, 0.80, 0.86, 1) was carried out: results are summarized in Fig. 2a for polycrystalline samples with ~60 Å thick $Ir_{1-x}Mn_x$ layers that are coupled to 60 Å thick Py films. The largest values of $\theta_{SH}^{eff}$ are found for x~0.53, 0.64 and 0.75. The two end members show much smaller values and, perhaps surprisingly, are of opposite sign (~+0.03 for Ir and ~-0.02 for Mn). The dependence of $\theta_{SH}^{eff}$ on the thickness of Ir, $Ir_{83}Mn_{17}$, $Ir_{47}Mn_{53}$, $Ir_{14}Mn_{86}$ and Mn, respectively, is shown in Fig. S1. These measurements indicate that the spin diffusion lengths in all of these metals are small (<~10 Å). We note that $\theta_{SH}^{eff}$ of these films exhibit little variation as a function of the RF excitation frequency (Fig. S2).

Epitaxial films of $IrMn_3$, oriented along the (100) and (111) facets, were prepared (see Methods) to explore the dependence of $\theta_{SH}^{eff}$ on the crystal orientation of the Ir-Mn. The crystal orientations were determined by θ-2θ X-ray diffraction (XRD) and diffractograms from selected areas of transmission electron micrographs, as shown in Fig. S3. Growth of $IrMn_3$ on (100) MgO and (0001) $Al_2O_3$ resulted in epitaxial $IrMn_3$ layers with (100) and (111) facets, respectively. Growth on amorphous $SiO_2$ on Si resulted in (111) textured polycrystalline $IrMn_3$ films. ST-FMR measurements as a function of RF frequency are summarized in Fig. S4. For (111) $IrMn_3$ and polycrystalline $IrMn_3$, $\theta_{SH}^{eff}$ exhibits little dependence on the RF frequency, while for (100) $IrMn_3$ $\theta_{SH}^{eff}$ is slightly higher at 9 and 10 GHz, which can be attributed to a larger error generated via the fit when $H_{res}$ is close to zero Oersted due to a stronger exchange bias field (Fig. S4). Note that no signal is detected, as expected, in devices formed from $IrMn_3$ films without any permalloy layers (Fig. S5).

Values of $\theta_{SH}^{eff}$ for (100) and (111) oriented IrMn$_3$/Py films are summarized and compared to results for polycrystalline IrMn$_3$/Py bilayers in Fig. 3a. In all cases $\theta_{SH}^{eff}$ depends weakly on the thickness of IrMn$_3$ for thicknesses in the range from ~30 to ~120 Å. A dramatic result is that the (100) oriented structures exhibit much larger $\theta_{SH}^{eff}$ ~0.20 than those exhibited either by p-IrMn$_3$ or (111) oriented IrMn$_3$. Moreover, as shown in Fig. 3b, a significant facet dependence is observed for x~75 and 80 atomic % but not for x~53 and 64% (see Fig. S5 for x-ray structural characterization).

The facet dependence of $\theta_{SH}^{eff}$ for IrMn$_3$ does not appear to be correlated with the resistivity of the films (see Table S1) nor with the magnetizations of the permalloy layers which are similar (see Fig. S7). Since the resistivity of the (100) oriented IrMn$_3$ is only 10-20% lower than that of (111) IrMn$_3$, it seems that it is very unlikely to account for the facet dependence of $\theta_{SH}^{eff}$ with interface transparency.

To explore the physical origin of the facet dependence of $\theta_{SH}^{eff}$ of IrMn$_3$ we consider the role of the in-plane exchange bias generated at the Py / IrMn$_3$ interface. Although H$_B$ of the as-deposited (100) IrMn$_3$/Py is significantly larger than that of (111) IrMn$_3$/Py (Figs. S8), the magnitude and direction of H$_B$ can readily be varied by annealing the bilayer in an in-plane magnetic field (±1 Tesla in vacuum). For the (100) 43 Å IrMn$_3$/Py the exchange bias field can be switched from an initial value of ~-300 Oe to ~+300 Oe by annealing at 150 °C for 30 min (Fig. 4a). Annealing at lower temperatures in fields that favor an opposite exchange bias yields intermediate or even zero H$_B$. As summarized in Fig. 4c, very little change in $\theta_{SH}^{eff}$ is observed as H$_B$ is varied widely by annealing. For the (111) IrMn$_3$/Py device, H$_B$ is significantly increased from a much smaller value in the as-deposited film to -120 Oe by annealing at temperatures up to 350 °C (Fig. 4b). Yet, again, as summarized in Fig. 5d, $\theta_{SH}^{eff}$ is hardly affected. Finally, $H_B$ exhibited by the polycrystalline $Ir_{1-x}Mn_x$ devices is very small for all Mn concentrations (after field annealing at 300 °C) except for x=0.64 ($H_B$=164 Oe) and 0.75 ($H_B$=37 Oe) (see Fig. 2b) yet $\theta_{SH}^{eff}$ is similar for x=0.53 ($H_B$ ~0) and x=0.64 and 0.75. These results clearly demonstrate no obvious correlation between $H_B$ and the magnitude of $\theta_{SH}^{eff}$. Perhaps this is not surprising since $H_B$ reflects the degree to which the magnetic moments within individual antiferromagnetic domains are uncompensated at the Ir-Mn/permalloy interface[21].

Notwithstanding the lack of correlation of $\theta_{SH}^{eff}$ with $H_B$, we find compelling evidence that the antiferromagnetic domain configuration of the $Ir_{1-x}Mn_x$ layer plays a key role in determining the magnitude of $\theta_{SH}^{eff}$ for x~0.75-0.80. In a first set of experiments, $IrMn_3$ based devices are annealed in a 1 T magnetic field at 300 °C that is aligned perpendicular, rather than parallel to the Py layers as in the experiments described above. This field is large enough to cause the magnetization of the Py layer to be oriented parallel to the field and thus perpendicular to the $IrMn_3$/Py interface at 300 °C. On cooling, the very large perpendicular exchange field imposed by the Py moments in direct contact with the $IrMn_3$ layer at the $IrMn_3$/Py interface thereby influences the configuration of the AF domains within the $IrMn_3$ layer. However, this mechanism is effective only when the blocking temperature of the $IrMn_3$ layer is below that of the anneal temperature, so that the moments within the AF domains can be rotated. Remarkably, we find that perpendicular field annealing results in a dramatic increase of $\theta_{SH}^{eff}$ by almost a factor of two to ~35 % for a (001) $IrMn_3$/Py device, when the $IrMn_3$ layer is ~3nm thick, as shown in Fig. 5a. By contrast, $\theta_{SH}^{eff}$ of the corresponding (111) device is hardly changed (Fig. 5b). Subsequently, annealing the same (001) oriented device in an in-plane magnetic field lowers $\theta_{SH}^{eff}$, but then annealing in a reverse perpendicular field (-1T) recovers the higher $\theta_{SH}^{eff}$ value (Fig. 5a). $\theta_{SH}^{eff}$ of the (111) oriented device is barely affected by these annealing procedures (Fig. 5b).

As shown in Fig. 5a, the increase in $\theta_{SH}^{eff}$ with perpendicular field annealing is strongly dependent on the thickness of the IrMn$_3$ (001) layer. When the IrMn$_3$ layer is too thin or too thick, no effect is found: an enhancement in $\theta_{SH}^{eff}$ is only found for intermediate IrMn$_3$ thicknesses in the range ~3-4 nm. This thickness dependence can be directly correlated with the thickness dependence of the blocking temperature (T$_B$) of the antiferromagnetic structure of the IrMn$_3$ layer. T$_B$, a temperature smaller than the Néel temperature, reflects the temperature below which there is sufficient anisotropy to freeze the magnetization of the AF domains[22-24]. The thickness dependence of $H_B$ and $H_C$ for (001) and (111) IrMn$_3$/Py at RT is shown in Fig. 5c. No exchange bias is found for 2 nm thick IrMn$_3$ layers consistent with it s T$_B$ (~200 K) being below the measurement temperature. Films that have a thickness ≥ 3nm exhibit an exchange bias field at RT, and, therefore T$_B$ > RT. The 8 nm thick film is sufficiently thick that T$_B$ is close to the bulk value[25] and is therefore much higher than the annealing temperature (300 °C). Thus, an important result is that $\theta_{SH}^{eff}$ is enhanced for perpendicular field annealing only when the AF structure of the IrMn$_3$ layer can be reconfigured by exchange coupling with the Py layer when it is cooled from a temperature above its blocking temperature and cooled through this temperature to "block" the AF domains at RT. Such a field cooling induced modification of the microscopic AF structure has previously been observed, for example, for AF NiO exchange coupled to ferromagnetic Co-Fe[26].

Note that when the exchange coupling between Py and IrMn$_3$ is weakened by the insertion of Cu or Au layers the Py layer will have less influence on the configuration of the AF domains in the IrMn$_3$ layer. Indeed, we find that $\theta_{SH}^{eff}$ decreases with increasing Au thickness for both (100) and (111) oriented devices to a facet independent value when the Au layer is ~10 Å (Fig. 4e and f). For Cu spacer layers $\theta_{SH}^{eff}$ decreases much more rapidly for (001) than for (111) oriented devices but a small difference in $\theta_{SH}^{eff}$ persists when the Cu layer is ~10 Å thick, likely due to the larger exchange interlayer coupling through Cu than through Au[27].

These results suggest that there are two distinct contributions to $\theta_{SH}^{eff}$: a first mechanism, which is facet independent, and which arises from bulk scattering within the IrMn$_3$ layer; and a second mechanism that is strongly facet dependent and which derives from the antiferromagnetic domains. IrMn$_3$ is known to display a triangular chiral magnetic structure with the Mn magnetic moments aligned at 120° to each other in the (111) plane[25,28] when the Ir and Mn are chemically ordered. Recently, it has been proposed that such structures will lead to an AHE even though they possess no net magnetization[15,16]. Using *ab-initio* calculations of the band structures and Berry curvatures (see Methods and supplementary information) we find that, in addition to an AHE, the triangular AF structure also gives rise to an intrinsic spin Hall conductivity (SHC) that is very large and, moreover, is strongly facet dependent (see Fig. 6 and Table 1). Furthermore, the calculated SHC agrees well with our experimental results. For a (111) IrMn$_3$ oriented film, an in-plane current generates spin currents inside the (111) plane but negligible spin currents in the out-of-plane direction, whereas, for the (001) film, the in-plane current leads to a large out-of-plane spin current whose amplitude is three times larger than that of the (111) in-plane case, as discussed in detail below.

In the cubic IrMn$_3$ lattice the Mn atoms are arranged in the form of triangles within the (111) plane of the primitive unit cell. Due to frustration, neighboring Mn moments align non-collinearly, at an angle of 120° to each other, to form two distinct AF arrangements, in which the Mn moments point towards (AF1) or out from the center of the triangle (AF2), respectively (see Fig.6a). AF1 and AF2 can be transformed into each other by a mirror reflection or a time reversal operation. We note that the mirror reflection, for example, with respect to the ($\bar{1}$10) plane, applies to both the lattice and magnetic moments, but that the time reversal operation only applies to the moments. Because AF1 and AF2 can be transformed from one to the other by a mirror symmetry then AF2 is the chiral counterpart of AF1. AF1 and AF2 will have the same energies and consequently can co-exist in the real material. It is known that under time reversal AHE is odd while SHE is even.[29] Thus it is clear from time reversal that AF1 and AF2 will exhibit the same SHC, but opposite AHC. This intuition is fully consistent with our *ab-initio* calculations, (taking into account the numerical inaccuracies), as shown in Figs. 6b and 6c. In addition, each Mn exhibits a moment of 2.91 μB from our calculations, which is consistent with previous work[15]. In addition, we calculate that there is a tiny out-of-plane net moment of 0.01 μB per Mn atom (along the [111] axis). This moment does not affect the values of SHC and AHC (as tested by setting this moment to zero and recalculating the results) but may be important in orienting the AF domains by perpendicular field annealing.

Our calculations show that the non-collinear AF order in the (111) plane generates a strong SHC that is highly anisotropic. The SHC ($\sigma_{ij}^k$; $i,j,k = x,y,z$) is a third-order tensor and represents the spin current $J_{S_i}^k$ generated by an electric field $E$ via $J_{S_i}^k = \sum_j \sigma_{ij}^k E_j$, where $J_{S_i}^k$ flows along the *i*-direction with the spin–polarization along the *k*–direction and $E_j$ is the *j*–component of $E$. Calculations of the SHC and AHC are shown in Table I for the two cases that correspond to our experiments where z is along (111) and (001), respectively. The magnetic structure of IrMn$_3$ belongs to the magnetic Laue group $\bar{3}m'1$, which thereby influences the SHC tensor by constraining some matrix elements to be equivalent or zero[30,31]. As shown in Table I, our calculated SHC matrices agree well with the symmetry-imposed tensor shape. Here we align the z–axis along the [111] crystallographic direction and xy–axes inside the (111) plane. The matrix elements $\sigma_{xy}^z = -\sigma_{yx}^z = -98\ (\hbar/e)$ (S/cm) indicate that an in-(111)-plane electric current can generate a considerable spin Hall current in the same plane. However, by contrast, the in-(111)-plane electric current generates a tiny spin current along the out of plane [111] direction: $\sigma_{zx}^y = -\sigma_{zy}^x = 5\ (\hbar/e)$ (S/cm).

We should point out that the shape of the SHC tensor depends on the coordinate system within which the directions of spin and currents are defined. If we rotate the z–axis from the [111] to the [001] direction, we obtain a different SHC tensor (see Table I). The two SHC tensors, however, can be readily transformed from one to the other by a corresponding rotation matrix (see supplementary information)[30,31]. The new coordinate system is convenient in interpreting the experimental results for the (001) oriented films. Here, the *E* field is along the x–axis ([110]) direction, and the spin current is along the z–axis. We obtain SHC $\sigma_{zx}^y = 315\ (\hbar/e)$ (S/cm), which is about three times larger in amplitude than that of the *in-plane* SHC ($\sigma_{xy}^z$) for the case where z is along [111].

The detailed values of SHC and AHC are shown in Table I for these two cases. These results show that for current along directions within the (111) plane there is negligible SHC perpendicular to this plane but that there is a fairly large SHC within the plane perpendicular to the charge current. On the contrary, for the (001) case when current is applied along the [110] direction within the (001) plane, a very significant SHC is generated along the direction perpendicular to this plane. These calculations are consistent with our speculation that the triangular AF structure generates a large SHC in the plane perpendicular to the chiral vector that defines the triangular lattice (i.e. perpendicular to the triangles defined by the Mn moments). And thus these calculations are consistent with a large spin-orbit torque for the (001) IrMn$_3$ samples that is generated by a novel SHC derived from the triangular AF structure, whereas no such torque is generated for the (111) oriented samples. Also, consistent with our perpendicular field annealing experiments, annealing in external magnetic fields of the opposite sign, we suppose, will switch the AF magnetic order between AF1 and AF2, but since the magnitude and sign of the SHC are not affected by the chirality of the AF structure the SHC is not affected. In our experiments, we find that perpendicular field annealing in fields of +1T and -1T has no effect on the magnitude of the measured spin-orbit torques (see Fig. 5a). By contrast, we anticipate that positive and negative magnetic field anneal treatments will change the sign of the AHC.

In summary, we have observed a giant $\theta_{SH}^{eff}$ in (100) oriented IrMn$_3$ and a smaller but still substantial $\theta_{SH}^{eff}$ for (111) oriented IrMn$_3$ films. $\theta_{SH}^{eff}$ is significantly enhanced by field cooling induced modifications of the microscopic antiferromagnetic structure for (001) oriented but not for (111) oriented IrMn$_3$/Py. We show from *ab-initio* calculations that these giant spin-orbit torques and their facet dependence arise from a novel spin Hall conductivity that is derived from the triangular chiral AF structure of IrMn$_3$. The discovery of a giant spin orbit torque at magnetic interfaces with antiferromagnetic metallic films that are already widely used in spintronics makes these films multi-functional and therefore of even greater interest for spintronics and spin-orbitronic applications.

## Methods

**Film growth and characterization**

All the films are grown in 3 mtorr argon in a magnetron sputtering system with a base pressure of ~ $1\times10^{-8}$ torr. The films of Ir, $Ir_{87}Mn_{13}$, $Ir_{47}Mn_{53}$, $Ir_{36}Mn_{64}$, $IrMn_3$, $Ir_{20}Mn_{80}$, $Ir_{14}Mn_{86}$, Mn, Py are grown from sputter targets of Ir, $Ir_{60}Mn_{40}$, $Ir_{36}Mn_{64}$, $Ir_{22}Mn_{78}$, $Ir_{18}Mn_{82}$, $Ir_{15}Mn_{85}$, $Ir_{11}Mn_{89}$, Mn and Py, respectively. TaN layer is grown by reactive sputtering of a Ta target in an argon - nitrogen gas mixture (ratio: 90/10). The concentration of Ir and Mn are determined by Rutherford backscattering spectrometry. Polycrystalline Ir-Mn films are grown on Si substrates (covered with 25 nm thick $SiO_2$). Single crystalline (100) and (111) $IrMn_3$ films are grown on (100) MgO and (0001) $Al_2O_3$ single crystalline substrates, respectively. The crystalline structure and orientation of the $IrMn_3$ is studied by θ-2θ X-ray diffraction (XRD), and by cross-section transmission electron microscopy. High resolution transmission electron microscopy (HRTEM) is carried out using a 200 kV JEOL 2010F field-emission microscope. The specimens were prepared using conventional cross-sectioning using mechanical dimpling followed by low energy ion milling while the sample is cooled.

**Device fabrication and measurement**

Devices for the ST-FMR measurements were fabricated using standard photo-lithography and argon ion milling procedures. In a first step rectangular micro-strips (100 μm long and 10 μm wide) are formed. In a second step two large electrical contact pads are formed from Ru/Au (5 nm and 50 nm). Electrical contacts are made using high-frequency contact probes. High frequency RF current (power = 14 dBm) is provided by a swept-signal generator (Agilent HP 83620B) and the DC mixed voltage is measured using a Keithley 2002 multimeter. Where not otherwise specified the ST-FMR measurements were carried out on devices that were not subject to any field anneal treatment.

*Ab-initio* **Calculations**

*Ab-initio* density-functional theory (DFT) calculations were performed to calculate the band structure of cubic $Mn_3Ir$ using the Vienna *ab-initio* simulation package (VASP)[32]. The generalized gradient approximation (GGA) was adopted to describe the exchange–correlation interactions between electrons in the Perdew–Burke–Ernzerhof (PBE) form[33]. We reproduced the in-plane non-collinear AFM configuration of Mn atoms and obtained the band structure in Fig. S10, which is consistent with previous theoretical and experimental reports[15,25,34]. Then we projected the DFT Bloch wave-functions onto the maximum localized Wannier functions by the Wannier90 package[35]. Based on the tight-binding Wannier Hamiltonian, we calculated the AHC and SHC. The AHC calculated by us is consistent with that in Ref. 15.

The intrinsic SHC was calculated by using the Kubo formula approach[36]:

$$\sigma_{ij}^k = -\frac{e\hbar}{V}\sum_{\vec{k}}\sum_n f_{\vec{k}n}\left(2\,\mathrm{Im}\sum_{n'\neq n}\frac{<u_{n\vec{k}}|\hat{J}_i^k|u_{n'\vec{k}}><u_{n'\vec{k}}|\hat{v}_j|u_{n\vec{k}}>}{\left(E_{n\vec{k}}-E_{n'\vec{k}}\right)^2}\right)$$

where the spin current operator $\hat{J}_i^k=\frac{1}{2}\{\hat{v}_i,\hat{s}_k\}$, the spin operator $\hat{s}_k$, the velocity operator $\hat{v}_i=\frac{1}{\hbar}\frac{\partial\hat{H}}{\partial k_i}$, $|u_{n\vec{k}}>$ is the eigen vector of Hamiltonian $\hat{H}$ and $i,j,k=x,y,z$. The SHC is a third-order tensor. The AHC and SHC are integrated in a k-point mesh of 101×101×101 in the first Brillouin zone. Using finer meshes of up to 201×201×201 only changes the results by less than 5%.

Contributions

W.Z. performed the device fabrication. W.Z. and W.H. performed the measurements and analyzed the data. S.H.Y. grew the films. Y.S, Y.Z. and B.Y carried out the *ab-initio* calculations. S.S.P.P. proposed and supervised the studies and postulated the SHC contribution from the triangular AF lattice. W.H., B.Y. and S.S.P.P. wrote the manuscript. All authors commented on the manuscript and contributed to its final version.

Acknowledgements

We gratefully acknowledge help from Chris Lada in designing the ST-FMR measurement system, Andrew Kellock in carrying out Rutherford backscattering spectrometry measurements, and Teya Topuria and Philip M. Rice for carrying out TEM measurements.

**Figure Captions**

**Figure 1 | ST-FMR measurement of $\theta_{SH}^{eff}$ in polycrystalline IrMn$_3$. a,** Illustration of ST-FMR experimental setup where $H_{ext}$ is the external field, $M$ is the magnetization of the permalloy layer (yellow), $\tau_H$ and $\tau_{ST}$ are the torques on M due to, respectively, the Oersted field created by the RF charge current, and the spin current in the Ir$_{1-x}$Mn$_x$ layer (green). **b.** ST-FMR spectra measured on a ~60 Å IrMn$_3$/~60 Å Py sample at a frequency varying from 9 to 12 GHz. **c.** ST-FMR spectrum measured at 9 GHz. The black solid lines are the fits with Lorentzian functions. The red and green solid lines represent the symmetric and antisymmetric Lorentzian fits. **d.** Frequency dependence of measured $\theta_{SH}^{eff}$ for ~60 Å IrMn$_3$/~60 Å Py. Inset: **e.** Spectra of ST-FMR measured on these devices with different thickness of IrMn$_3$ at 11 GHz. **f.** Thickness dependence of measured $\theta_{SH}^{eff}$ for IrMn$_3$/~60 Å Py. Error bars correspond to one standard deviation in **d** and **f**.

**Figure 2 | Dependence of SHA and $H_B$ on Mn concentration in Ir$_{1-x}$Mn$_x$/ Py bilayers. a,** $\theta_{SH}^{eff}$ measured at 11 GHz, and **b,** in-plane exchange bias field for polycrystalline bilayers of ~60 Å Ir$_{1-x}$Mn$_x$ / ~60 Å Py versus Mn concentration x. Note that $H_B$ was measured after the films were annealed in an in- plane field of 1T for 30 min at 300 °C.

**Figure 3 | Facet dependent $\theta_{SH}^{eff}$ in crystalline Ir$_{1-x}$Mn$_x$. a,** $\theta_{SH}^{eff}$ as a function of IrMn$_3$ thickness for (100) (olive circles), (111) (blue circles) and polycrystalline (cyan circles) oriented films, respectively. **b,** $\theta_{SH}^{eff}$ as a function of Mn concentration for (001), (111) and polycrystalline Ir$_{1-x}$Mn$_x$. Error bars correspond to one standard deviation in **a** and **b**.

**Figure 4 | a-d,** In-plane exchange bias field and $\theta_{SH}^{eff}$ as a function of annealing temperature for (100)-IrMn$_3$ (**a, c**), (111)-IrMn$_3$ (**b, d**), respectively. The thickness of IrMn$_3$ is 43 Å. Squares and circles indicate positive and negative in-plane magnetic field (1T) during annealing (30 min), respectively. **e**, **f,** $\theta_{SH}^{eff}$ as a function of Cu and Au insertion layers, respectively, for (100) and (111) oriented IrMn$_3$/Py bilayers (structure shown in insets). Error bars correspond to one standard deviation in **e** and **f**.

**Figure 5 |** $\theta_{SH}^{eff}$ for **a** (100) and **b** (111) IrMn$_3$/ 60 Å Py bilayers as a function of IrMn$_3$ thickness measured sequentially in: as grown; annealed in 1 T perpendicular field (60 min); annealed in an in-plane 1T field (30 min); and annealed in -1 T perpendicular field (60 min) devices. **c,** exchange bias field (solid symbols) and coercive field (open symbols) measured in (100) and (111) IrMn$_3$/ 60 Å Py bilayers as a function of IrMn$_3$ thickness after the unpatterned films were annealed in-plane in a 1T field (30 min) using vibrating sample magnetometry. The annealing field was aligned along the (001) axis for the (001) oriented samples. The exchange bias field was also measured in the same devices used in **a** and **b** via anisotropic magnetoresistance measurements (see Fig. S8).

**Figure 6. Chiral noncollinear antiferromagnetic (AFM) orders and corresponding anomalous Hall conductivity (AHC) and spin Hall conductivity (SHC). a**, Schematic diagram of two chiral AF orderings, AF1 and AF2, of the Mn moments in the (111) plane of IrMn$_3$. **b,c** *Ab-initio* calculations of the AHC ($\sigma_{xy}$) and SHC ($\sigma_{xy}^z$) with respect to chemical potential. The charge-neutral point is set to zero. Here the *x–, y–* and *z–*axes correspond to the crystallographic directions [1$\bar{1}$0], [$\bar{1}\bar{1}$2] and [111], respectively. Because AF1 and AF2 can be

transformed into each other by a mirror reflection ($\mathcal{M}$) or a time reversal ($\mathcal{T}$) operation, they exhibit opposite signs of AHC but the same sign of SHC.

Table I. **The AHC and SHC tensors**. The magnetic lattice of IrMn can be classified into the magnetic Laue group $\bar{3}m'1$, which determines the symmetry and shape of the tensors (second row). The calculated results of AHC and SHC (third row) agree well with the symmetry-imposed tensor shape. We only show results for the AF1 order, since those of AF2 can be obtained simply by the symmetry analysis. When rotating the $z$–axis from the crystallographic [111] direction to the [001] direction, the tensor matrices are transformed into a new shape (the fourth row). The AHC is in the unit of S/cm and SHC is in the unit of $(\hbar/e)(\text{S/cm})$.

Figure 1

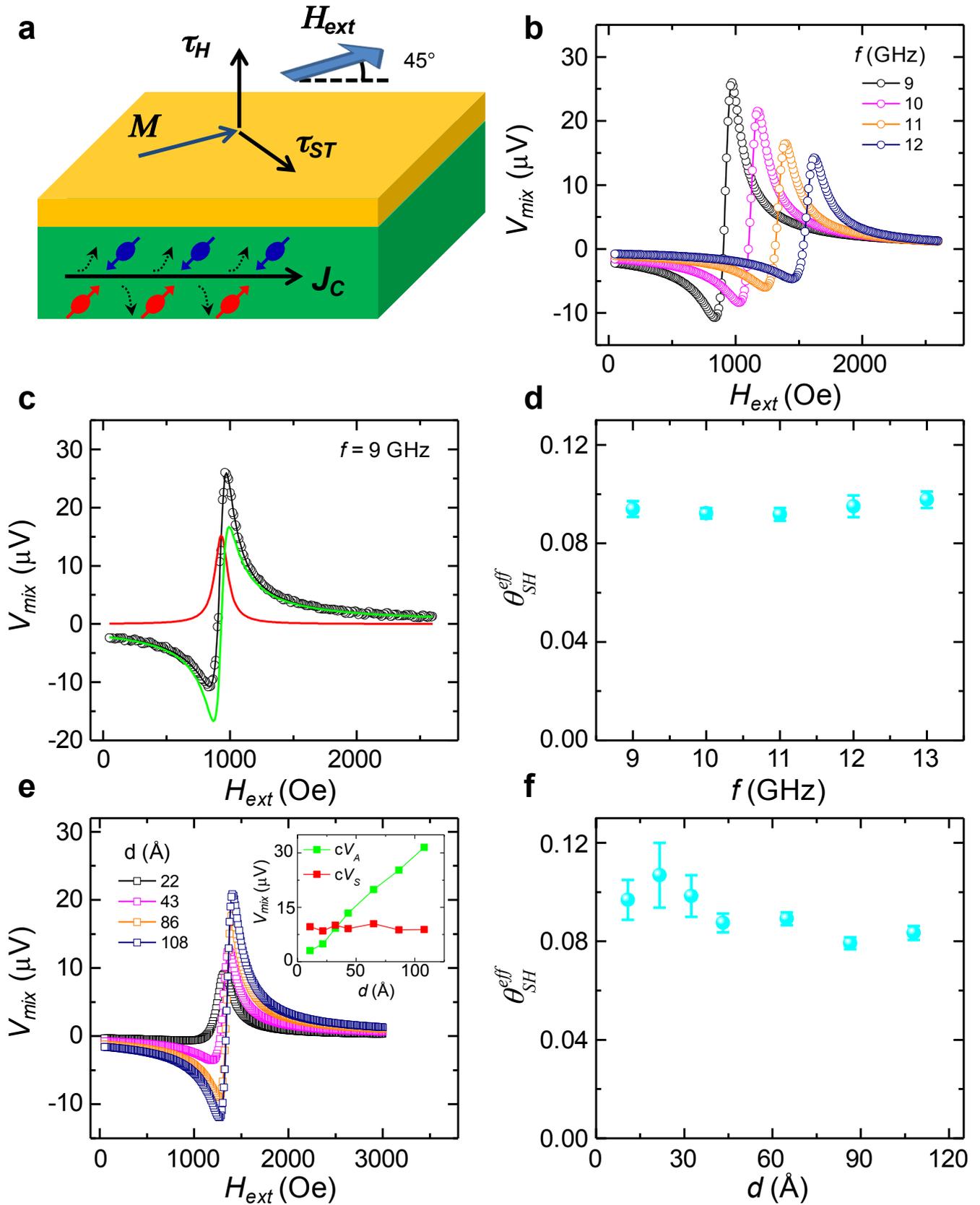

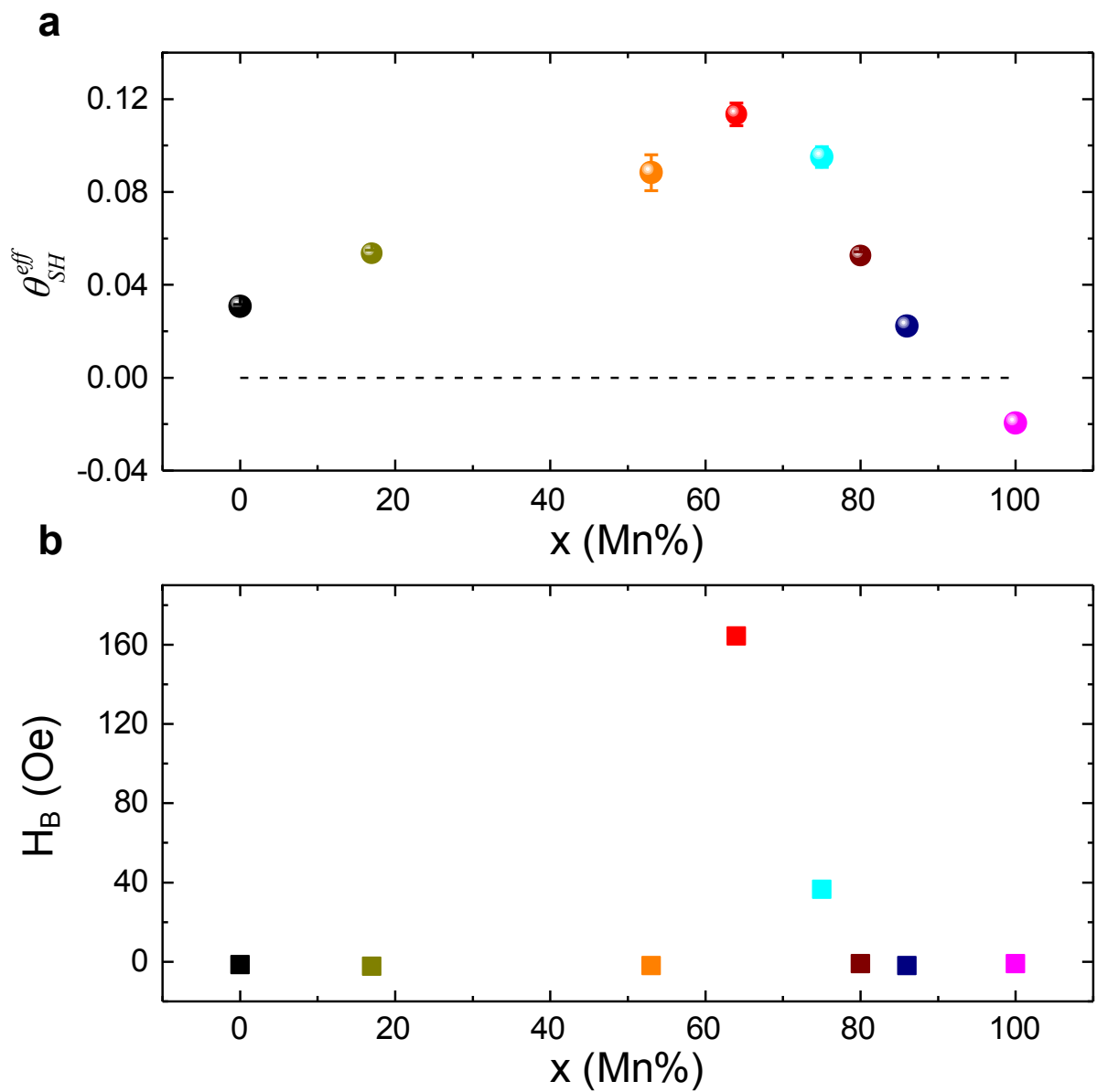

Figure 3

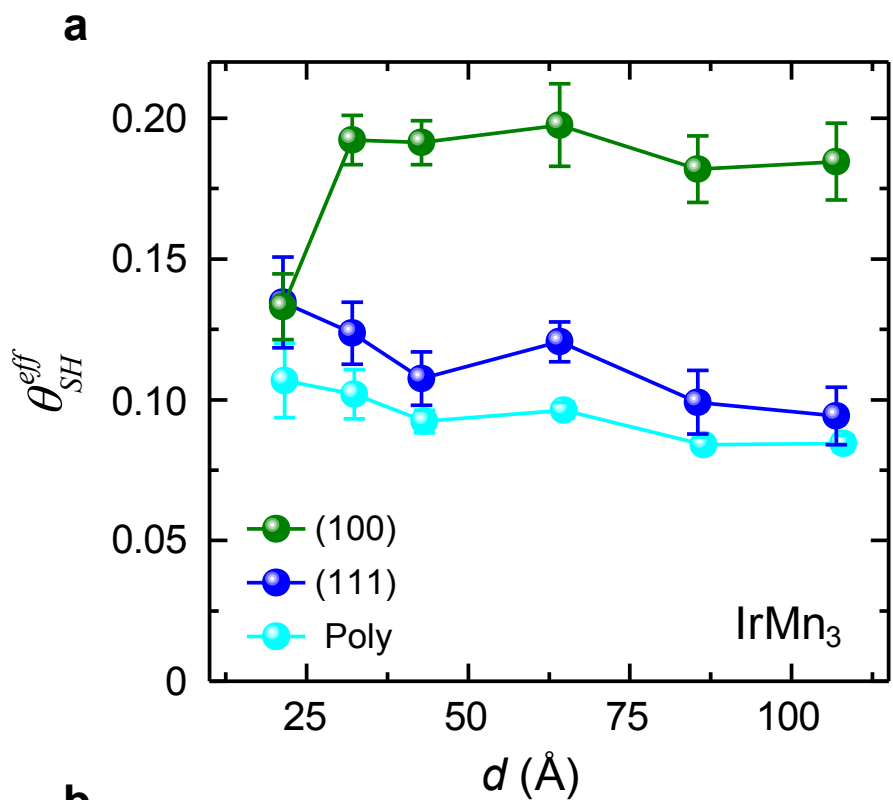

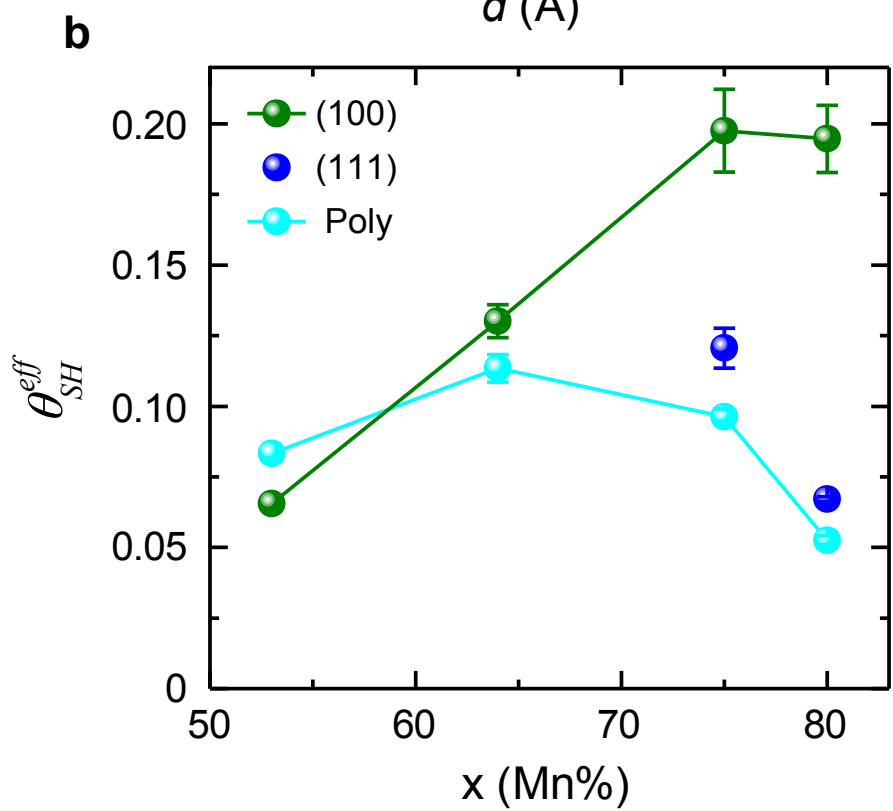

Figure 4

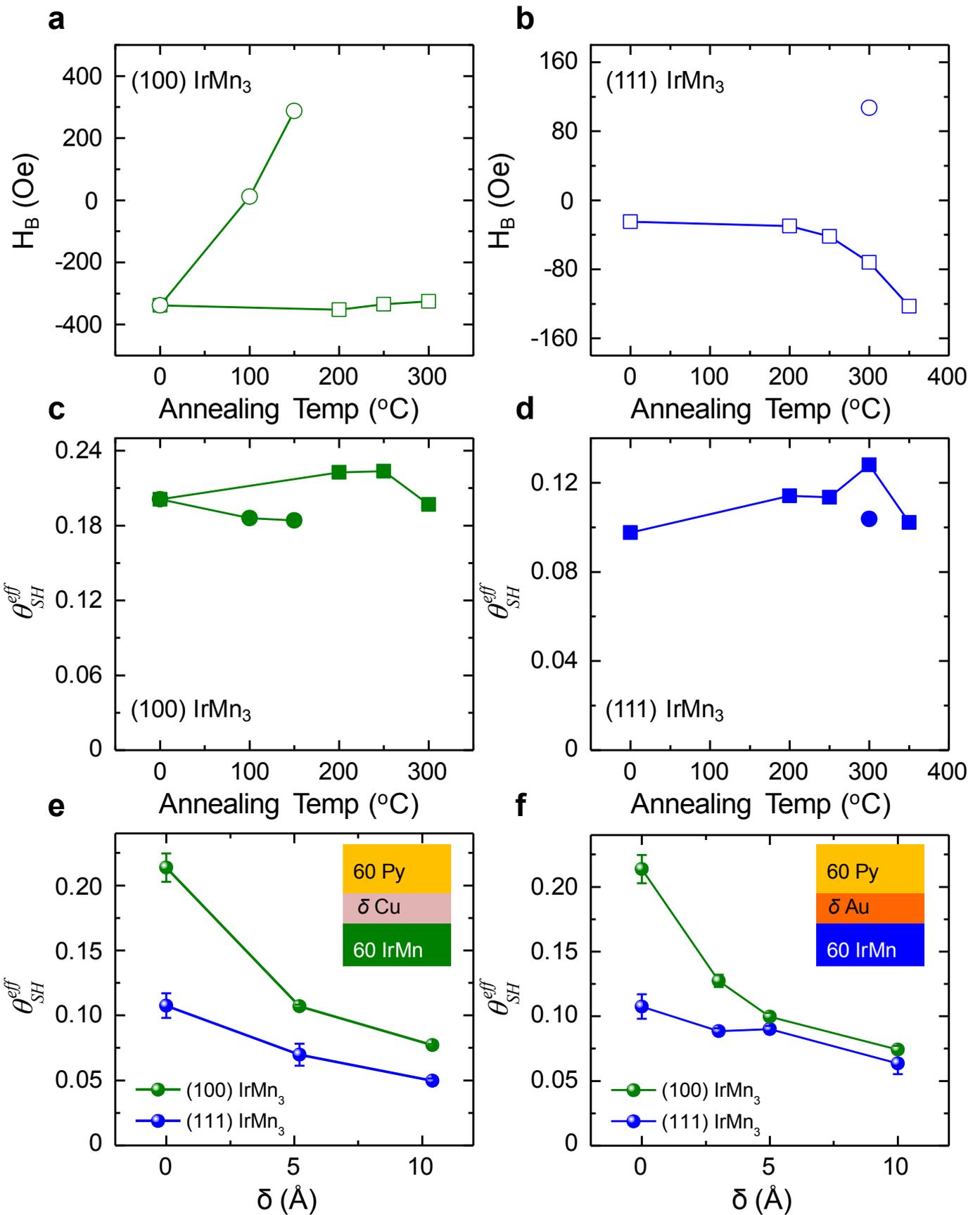

Figure 5

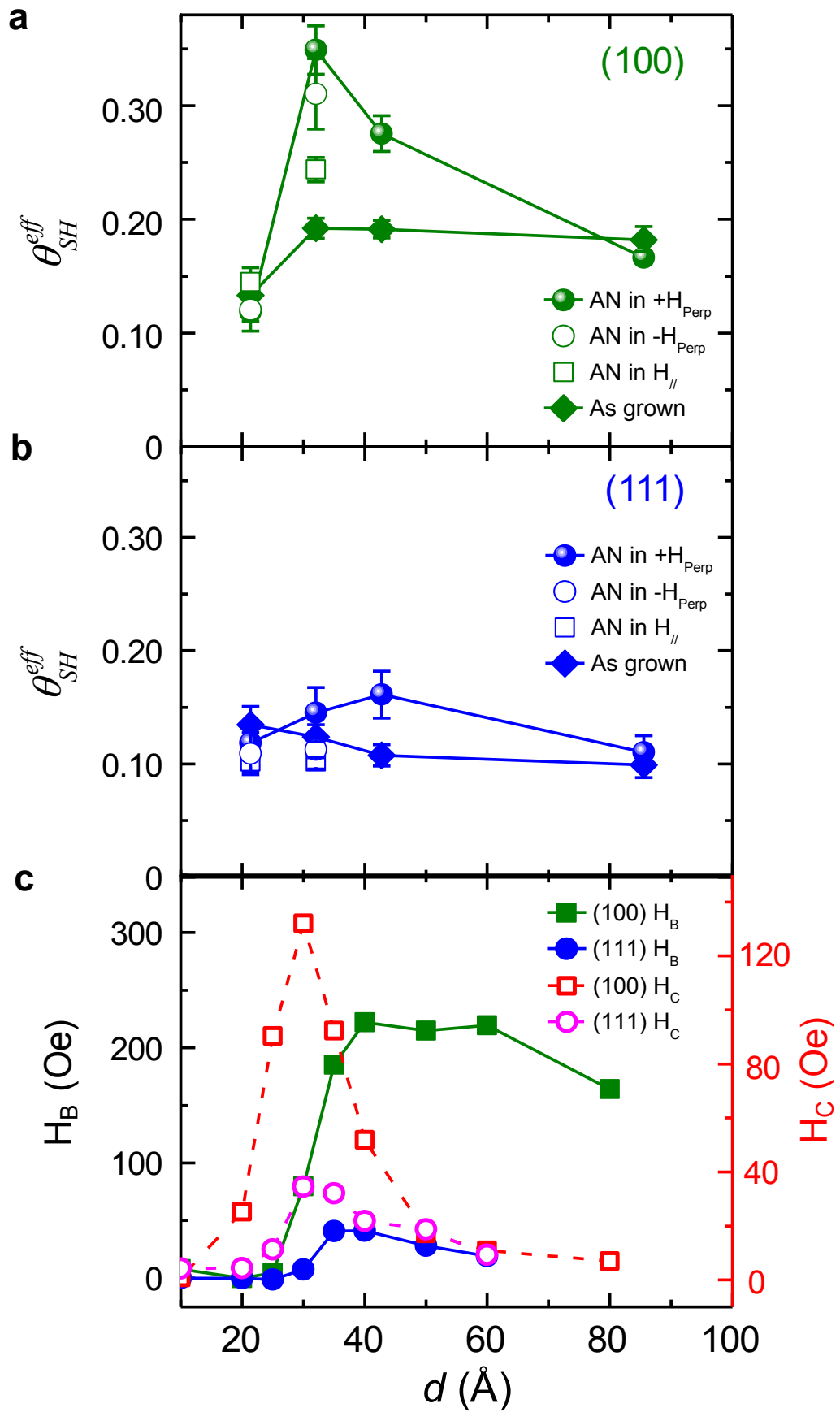

Figure 6

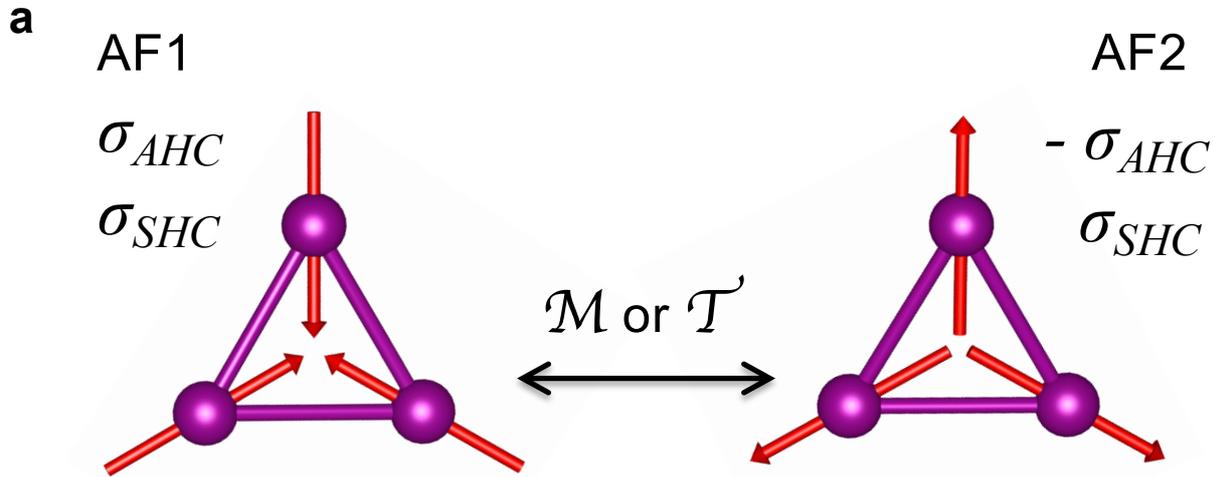
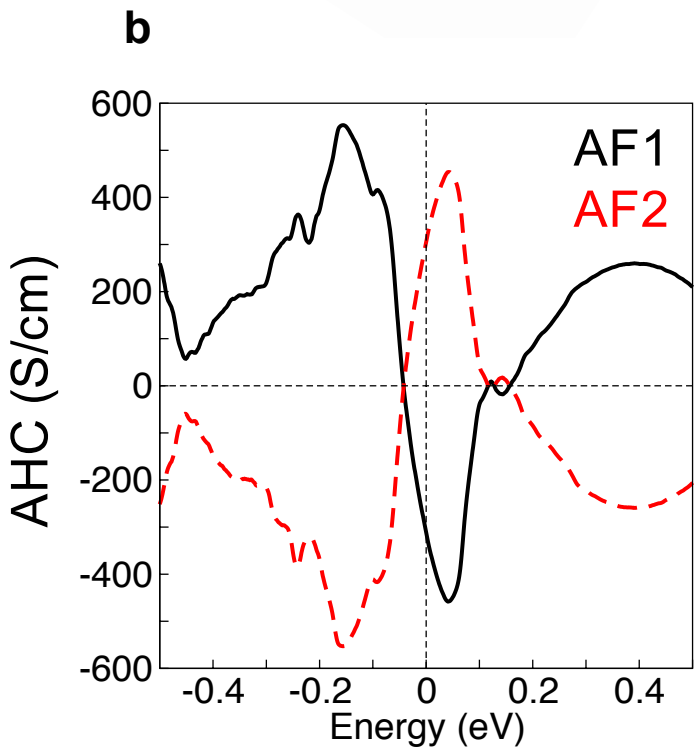
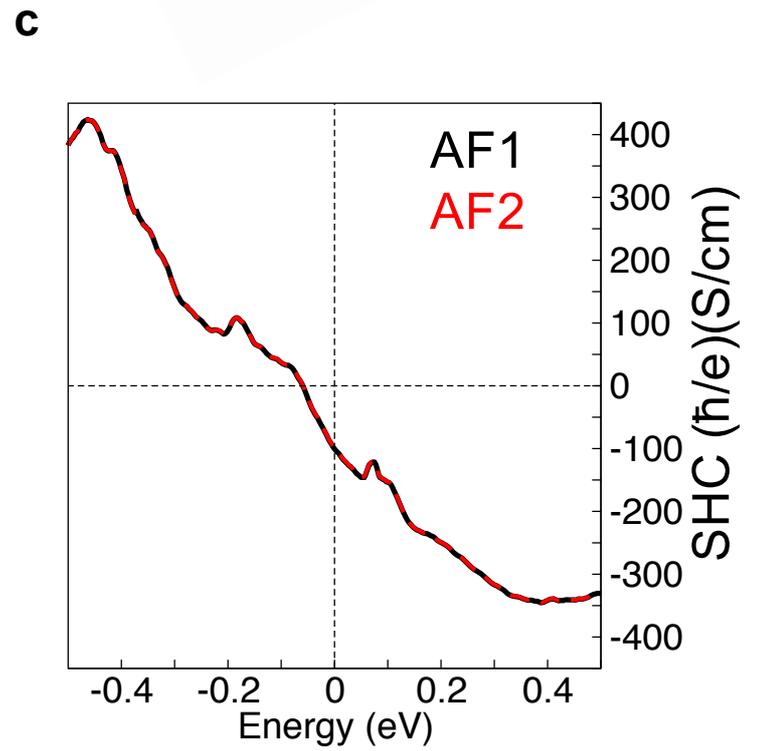

| | AHC | SHC | | |
|---|---|---|---|---|
| | $\sigma$ | $\underline{\sigma}^x$ | $\underline{\sigma}^y$ | $\underline{\sigma}^z$ |
| Magnetic Laue group $\bar{3}m'1$ $z$ [111] | $\begin{pmatrix} 0 & \sigma_{xy} & 0 \\ -\sigma_{xy} & 0 & 0 \\ 0 & 0 & 0 \end{pmatrix}$ | $\begin{pmatrix} \sigma^x_{xx} & \sigma^y_{xx} & \sigma^x_{xz} \\ \sigma^y_{xx} & -\sigma^x_{xx} & -\sigma^y_{xz} \\ \sigma^x_{zx} & -\sigma^y_{zx} & 0 \end{pmatrix}$ | $\begin{pmatrix} \sigma^y_{xx} & -\sigma^x_{xx} & \sigma^y_{xz} \\ -\sigma^x_{xx} & -\sigma^y_{xx} & \sigma^x_{xz} \\ \sigma^y_{zx} & \sigma^x_{zx} & 0 \end{pmatrix}$ | $\begin{pmatrix} \sigma^z_{xx} & \sigma^z_{xy} & 0 \\ -\sigma^z_{xy} & \sigma^z_{xx} & 0 \\ 0 & 0 & \sigma^z_{zz} \end{pmatrix}$ |
| Calculation $z$ [111], $x$ [1$\bar{1}$0], $y$ [$\bar{1}\bar{1}$2] | $\begin{pmatrix} 0 & -312 & 0 \\ 312 & 0 & 0 \\ 0 & 0 & 0 \end{pmatrix}$ | $\begin{pmatrix} 216 & 0 & 0 \\ 0 & -216 & 318 \\ 0 & -5 & 0 \end{pmatrix}$ | $\begin{pmatrix} 0 & 216 & -318 \\ 216 & 0 & 0 \\ 5 & 0 & 0 \end{pmatrix}$ | $\begin{pmatrix} 0 & 98 & 0 \\ -98 & 0 & 0 \\ 0 & 0 & 0 \end{pmatrix}$ |
| Calculation $z$ [001], $x$ [110], $y$ [$\bar{1}$10] | $\begin{pmatrix} 0 & -180 & 0 \\ 180 & 0 & -255 \\ 0 & 255 & 0 \end{pmatrix}$ | $\begin{pmatrix} 0 & -121 & 0 \\ 124 & 0 & 142 \\ 0 & 165 & 0 \end{pmatrix}$ | $\begin{pmatrix} -220 & 0 & -8 \\ 0 & 216 & 0 \\ 315 & 0 & 4 \end{pmatrix}$ | $\begin{pmatrix} 0 & 72 & 0 \\ -78 & 0 & -340 \\ 0 & -95 & 0 \end{pmatrix}$ |